\documentclass[twocolumn,showpacs,amsmath,amssymb,prb]{revtex4-2}

\usepackage{natbib}

\usepackage[version=4]{mhchem}
\usepackage{physics}
\usepackage{siunitx}
\usepackage{graphicx}
\usepackage[table]{xcolor}
\usepackage{lipsum}
\usepackage[T1]{fontenc}
\usepackage[format=hang,justification=centerlast,font={small,stretch=1},labelfont=bf,labelsep=colon]{caption}
\usepackage{subcaption}
\usepackage[backref=true,colorlinks=true,citecolor=blue,urlcolor=blue,linkcolor=blue]{hyperref}

\begin{document}

\title{Impact of a ferromagnetic insulating barrier in magnetic tunnel junctions}
\author{M. Abbasi Eskandari}
\author{S. Ghotb}
\author{P. Fournier}
\affiliation{Institut quantique, Regroupement québécois sur les matériaux de pointe et Département de physique, Université de Sherbrooke, Sherbrooke, J1K 2R1, Québec, Canada}

\begin{abstract}
We investigate spin-dependent conductance across a magnetic tunnel junction (MTJ) including a ferromagnetic insulating barrier. The MTJ consists of two half-metallic ferromagnetic \ce{La_{2/3}Sr_{1/3}MnO_{3}} (\ce{LSMO}) manganites as electrodes and \ce{La_{2}NiMnO_{6}} (\ce{LNMO}) double perovskite as a ferromagnetic insulating barrier. The resistance of the junction is strongly dependent not only on the orientation of the magnetic moments in \ce{LSMO} electrodes, but also on the direction of the magnetization of the \ce{LNMO} barrier with respect to that of \ce{LSMO}. The ratio of tunnel magnetoresistance reaches a maximum value of \SI{24}{\%} at \SI{10}{\kelvin}, and it decreases with temperature until it completely disappears above the critical temperature of \ce{LNMO} at \SI{280}{\kelvin}. The tunneling process is described using a mechanism which involves both empty and filled $e_{g}$ states of the \ce{LNMO} barrier acting as a spin-filter. A magnetic insulating barrier is an interesting path for achieving room temperature magnetoresistance in oxide-based heterostructures. 
\end{abstract}

\maketitle

\section{Introduction}

Spintronics uses the electron spin degrees of freedom to manipulate the electron transport or to store information. An entirely new generation of electronic devices has emerged with spintronics featuring non-volatile storage, ultra-fast switching, reduced energy consumption and increased integration density \cite{wolf2001spintronics, bratkovsky2008spintronic}. Tunnel magnetoresistance (TMR), one of the most important phenomena in spintronics, was first discovered by Julli\`ere in \num{1975} in magnetic tunnel junctions (MTJs) \cite{julliere1975tunneling}. Since then, MTJs have generated considerable interest due to their potential applications in spin-electronic devices such as magnetic sensors and magnetic random-access memories (MRAMs) \cite{maekawa2002spin}. MTJs consist of two ferromagnetic metallic layers separated by a thin insulating barrier. In MTJs, the ratio of TMR depends on the relative orientation of the magnetization in the two ferromagnetic layers on each side of the barrier, which can be controlled by an external magnetic field. The tunnel magnetoresistance can be expressed in term of the junction resistances when the magnetic moments of two ferromagnets are parallel ($R_{\mathrm{P}}$) and antiparallel ($R_{\mathrm{AP}}$), as follows: \[TMR =(R_{\mathrm{AP}}-R_{\mathrm{P}})/R_{\mathrm{P}} \times 100\]

It has been shown that half-metallic materials that possess only one spin polarization at the Fermi level can produce a very large TMR ratio due to their large spin polarization \cite{bratkovsky1997tunneling}. Among the materials that are half-metals, manganites are considered as popular choices for ferromagnetic electrodes in MTJs due to their large spin polarization and tunable transition temperature that can be used to design electronic devices to meet specific functional requirements. The best results on manganite-based MTJs have been reported with optimally-doped \ce{La_{2/3}Sr_{1/3}MnO_{3}} (\ce{LSMO}) manganite \cite{lu1996large, sun1997temperature, viret1997low}, particularly it showed a TMR as high as \SI{1850}{\percent} in a \ce{LSMO/SrTiO_{3}/LSMO} MTJ, corresponding to a spin polarization of \SI{95}{\percent} for \ce{LSMO} at low temperature \cite{bowen2003nearly}.

Two types of barrier have been extensively studied in MTJs including amorphous and crystalline insulating barriers. Amorphous barriers such as \ce{AlO_{x}} were a common choice in the first generation of MTJs due to their ease of fabrication process, spin conservation across the barrier and pinhole-free layers \cite{meservey1994spin, moodera1995large}. MTJs with amorphous barrier never showed TMR larger than \SI{81}{\percent} at room temperature which were in close agreement with Julli\`ere's model prediction \cite{wei200780}. The most remarkable results were obtained using crystalline insulating compounds such as \ce{MgO} as the tunnel barrier, leading to a large TMR of up to \SI{600}{\percent} at room temperature due to coherent tunneling through the barrier \cite{bowen2001large, ikeda2008tunnel}. Since then, crystalline insulating barriers have been the focus of MTJs' studies.

Due to the rarity of ferromagnetic insulators, this type of barriers has not been explored extensively. Only a few studies have investigated the effect of magnetic barriers on TMR in spin-filter junctions \cite{gajek2005spin, luders2006spin, leclair2002large}. For instance, Gajek \textit{et al} \cite{gajek2005spin} have demonstrated that a TMR of up to \SI{50}{\percent} can be obtained in \ce{Au/BiMnO_{3}/LSMO} junctions according to whether the magnetization of \ce{BMO} and \ce{LSMO} are parallel or opposite. It should also be noted that spin-filter junctions usually operates at low temperature and the TMR decays very fast with temperature.

In the present work, we use \ce{La_{2}NiMnO_{6}} (\ce{LNMO}) double perovskite to explore the impact of a ferromagnetic barrier on the TMR of MTJs. The insulating nature of \ce{LNMO} provides the tunneling conditions in the entire temperature range of operation, combined with a ferromagnetic order with a transition temperature ranging from \SIrange[range-units=single]{180}{285}{\kelvin} depending on the level of cationic ordering in the sample. We also employ half-metallic \ce{LSMO} manganite as the electrodes to maximize spin polarization. We explore how the tunneling occurs through the magnetic barrier and propose a mechanism involving empty and filled $e_{g}$ states around the Fermi level in the barrier. It is also shown that the device can be operated up to \SI{280}{\kelvin}, close to the maximum magnetic phase transition temperature of the \ce{LNMO} barrier.

\section{Experiments and methods}

\begin{figure}
\center
\includegraphics[scale=0.35]{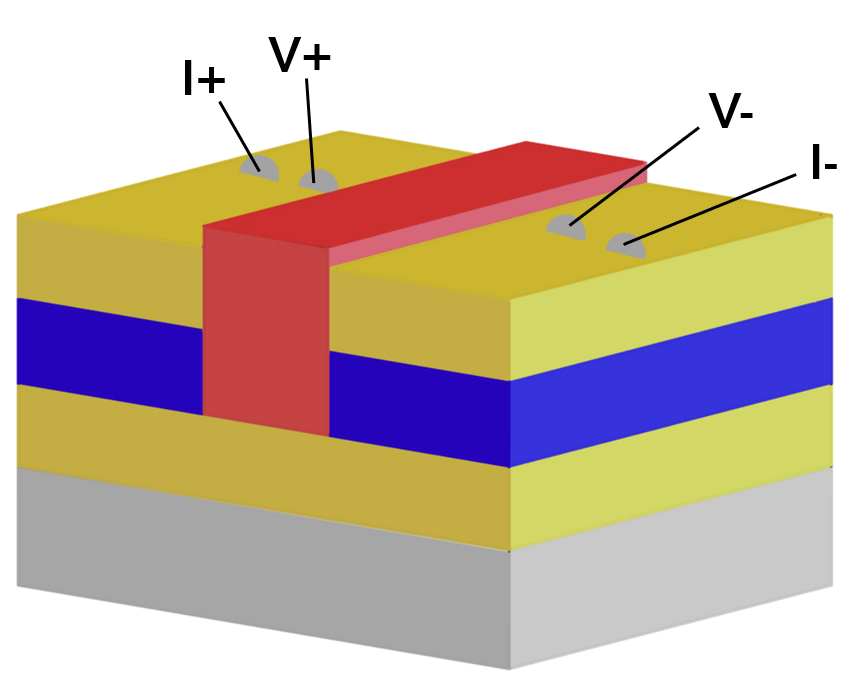}
\caption{Schematic of the MTJ device consisting of two \ce{La_{2/3}Sr_{1/3}MnO_{3}} electrodes (yellow) separated by a \ce{La_{2}NiMnO_{6}} barrier (blue). The middle slab (red) separating two devices is made of an amorphous layer of \ce{SiO_{2}} deposited on the bottom layer of epitaxial \ce{LSMO}. This slab includes also an insulating non-epitaxial \ce{LNMO/LSMO} cover.}
\label{device}
\end{figure}

The MTJ devices consist of a ferromagnetic insulating \ce{LNMO} barrier sandwiched between two  half-metallic \ce{LSMO} layers as the electrodes. Pulsed laser deposition (PLD) technique has been used to grow epitaxial layers of \ce{LSMO} and \ce{LNMO} on $(001)$-oriented \ce{SrTiO_{3}} substrates. A schematic illustration of the final device is displayed in Figure~\ref{device}. In order to fabricate a MTJ device, first a layer of \ce{LSMO} is deposited on the substrate, followed by patterning a resin layer using photographically technique defining a strip of exposed \ce{LSMO} in the middle of the layer. Then, a \SI{100}{\nm}-thick amorphous \ce{SiO_{2}} layer is deposited using RF magnetron sputtering. Following a lift off that removes the resin covered by the amorphous layer and leaves a strip of \ce{SiO_{2}} on top of \ce{LSMO}, a \ce{LNMO} layer and then a \ce{LSMO} layer are deposited. The amorphous strip of \ce{SiO_{2}} prevents the epitaxial growth of \ce{LNMO} and \ce{LSMO} layers on top of it, imposing a current flow through the barrier with the contact configuration shown in Fig.~\ref{device}. The bottom and top \ce{LSMO} electrodes are \SI{50}{\nm} thick, while the thickness of \ce{LNMO} barrier is \SI{40}{\nm}. Also, the surface area of each junction is around \SI{0.5}{\mm^{2}}. Structural characterization were performed using a Bruker AXS D$8$-diffractometer with $\ce{CuK_{\alpha 1}}$ radiation in $2\theta / \omega$ configuration. Further surface investigations have been performed using a Veeco Dimension Icon Atomic Force Microscope (AFM). A physical properties measurement system (PPMS) from Quantum Design was employed to carry out the transport measurements with the help of a horizontal rotator option allowing to apply magnetic field in different directions with respect to the interfaces of the sample, in the temperature range of \SIrange{10}{300}{\kelvin}. Finally, the magnetization measurements were performed using the reciprocating sample option (RSO) of a \SI{7}{\tesla} SQUID magnetometer from Quantum Design.

\section{Results and discussion}

\begin{figure}
\center
\includegraphics[scale=0.4]{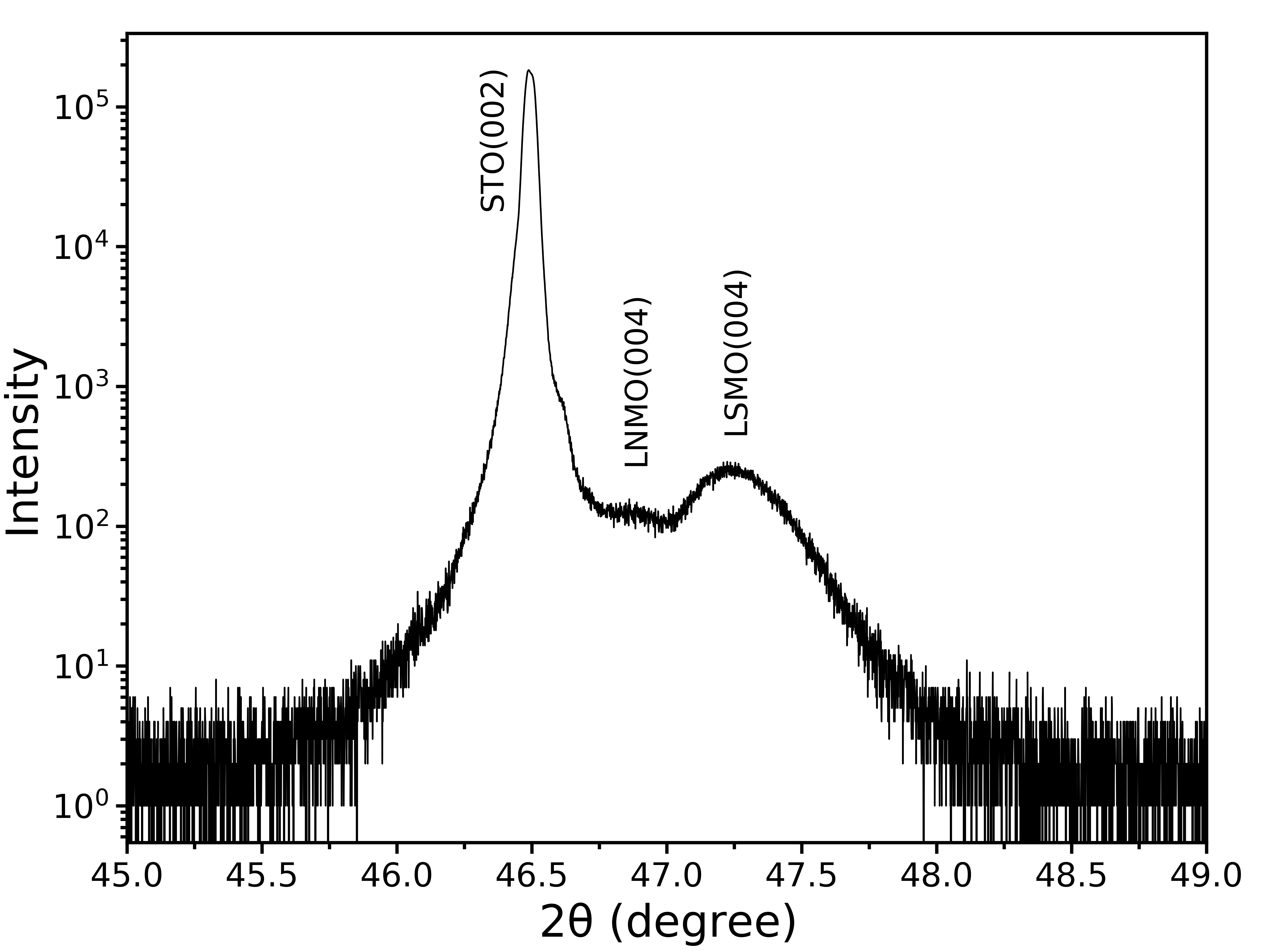}
\caption{X-ray diffraction (XRD) pattern of a trilayer MTJ device consisting of two \ce{LSMO layers} and one \ce{LNMO} layer on a \ce{STO} substrate.}
\label{xrd}
\end{figure}

The XRD $2\theta /\omega$ scan of the sample in the range from \ang{10} to \ang{80} (see supplemental material) confirms the absence of impurity or secondary phases in the samples by assigning all the peaks to \ce{LSMO} and \ce{LNMO} layers. A magnified view of a XRD pattern around the $(002)$ peak from the substrate is displayed in Figure~\ref{xrd} where the $(004)$ reflections from the \ce{LSMO} and \ce{LNMO} layers can be clearly seen, indicating the out-of-plane growth of these epitaxial layers. Furthermore, AFM measurements show a surface roughness less than \SI{1}{\nm} over a lateral distance of \SI{5}{\mu m} for a \SI{50}{\nm}-thick \ce{LSMO} layer (see supplemental material), confirming the sharp and smooth interfaces between the layers distributed only at the unit cell step level.

\begin{figure}
\center
\includegraphics[scale=0.45]{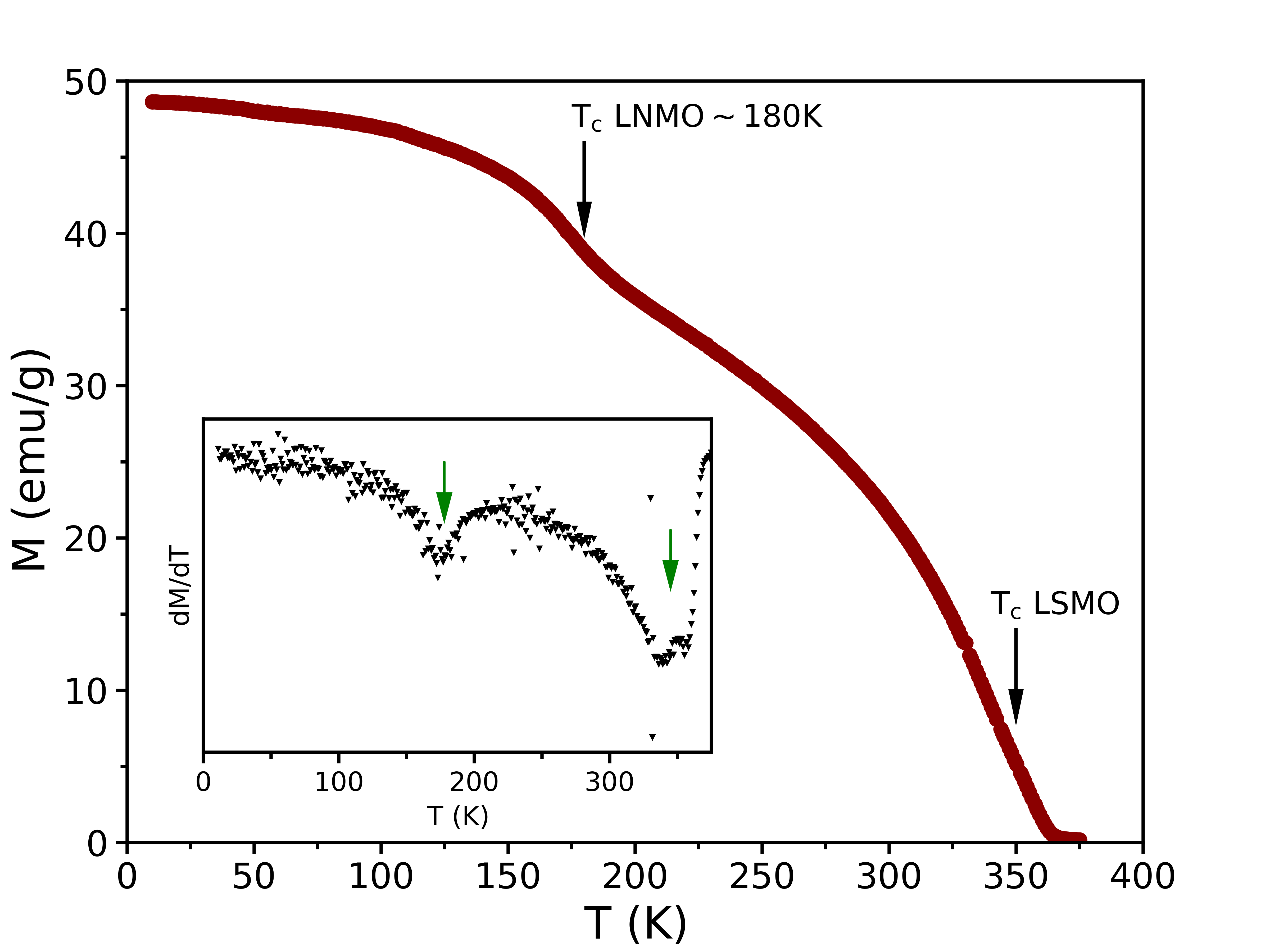}
\caption{Field-cooled magnetization as a function of temperature under a magnetic field of \SI{200}{Oe} for a \ce{LSMO/LNMO/LSMO} MTJ device. The inset shows the derivative of the magnetization versus temperature.}
\label{MvsT}
\end{figure}

Field-cooled magnetization of a MTJ device was measured as a function of temperature at a fixed magnetic field of \SI{200}{Oe} in the temperature range from \SIrange[range-units = single]{10}{370}{\kelvin}. As depicted in Figure~\ref{MvsT}, the sample clearly goes through two magnetic phase transitions at approximately \SIlist[list-units = single]{180;350}{\kelvin}, corresponding to the ferromagnetic-paramagnetic transitions of \ce{LNMO} and \ce{LSMO} layers, respectively. The transition temperatures were determined from the minimum in the derivative of the magnetization with respect to temperature (inset of Fig.~\ref{MvsT}). The low transition temperature of the \ce{LNMO} layer compared to its maximum $T_{\mathrm{c}}$ of \SIrange[range-phrase = --]{280}{285}{\kelvin} \cite{kitamura2009ferromagnetic} is ascribed to a low cation-ordering level in the system, where \ce{Mn^{4+}} and \ce{Ni^{2+}} ions occupy $\ce{B/B^{\prime}}$-sites randomly with partial ordering.

\begin{figure}
\center
\includegraphics[scale=0.45]{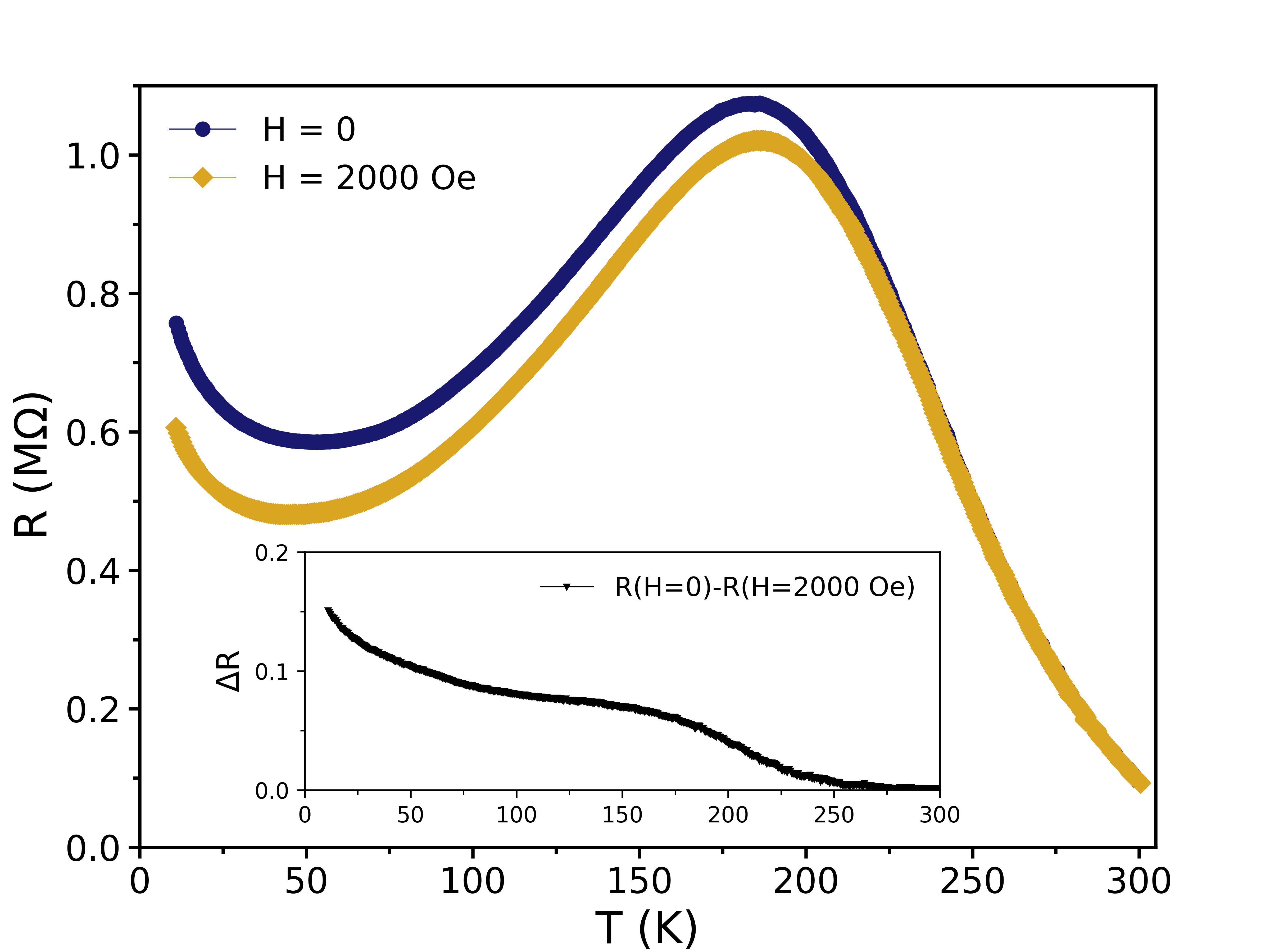}
\caption{Temperature dependence of the junction resistance at zero and \SI{0.2}{\tesla} magnetic field. Inset displays the difference between the two curves (magnetoresistance) which goes to zero at \SI{275}{\kelvin}.}
\label{RvsT}
\end{figure}

Figure~\ref{RvsT} shows the temperature dependence of the junction resistance under zero and \SI{0.2}{\tesla} magnetic field applied parallel to the surface of a trilayer device. Similar to typical oxide-based MTJs, our devices exhibit different behaviors in different temperature regions. In the high temperature region from \SIrange[range-units = single]{185}{300}{\kelvin}, the junction shows a semiconducting-like behavior where the resistance increases with decreasing temperature confirming direct tunneling transport in the junction \cite{jonsson2000reliability}. This is a common characteristic feature in all MTJs, regardless of their compositions \cite{bowen2001large, xia2021angular, sun1997temperature}. However, below \SI{185}{\kelvin}, the junction enters a low temperature region in which the resistance decreases with decreasing temperature. There have only been a few oxide-based MTJs that exhibit this unusual metallic-like resistance. Its origin has not yet been determined, but it was proposed that it may be the result of oxygen deficiencies at the interface when the barrier is an insulator \cite{sun1997temperature, galceran2016tunneling, viret1997low}. In our case, this anomaly in our MTJ devices could instead be attributed to the onset of magnetic order around \SI{200}{\kelvin} in the \ce{LNMO} barrier as seen in Fig.~\ref{MvsT}. The decreasing resistance would then imply an increase of the tunneling probability through the \ce{LNMO} barrier as its magnetization grows. The application of \SI{0.2}{\tesla} magnetic field lowers further the junction resistance in the same temperature range below \SI{200}{\kelvin}. In fact, the observed magnetoresistance goes to zero at \SI{275}{\kelvin} as shown in the inset of Fig.~\ref{RvsT}. This onset temperature is very close to the maximum $T_{\mathrm{c}}$ of \SI{285}{\kelvin} observed for cation-ordered \ce{LNMO} \cite{kitamura2009ferromagnetic}. Although we cannot really observe a magnetic transition around \SIrange[range-phrase=--]{275}{285}{\kelvin} in the M(T) curve in Fig.~\ref{MvsT}, we cannot rule out the presence of domains with a high degree of cationic ordering with such high $T_{\mathrm{c}}$. Moreover, in contrast with the typical behavior of colossal magnetoresistance in manganites which usually shows a shift in the resistance peak to higher temperatures under an applied magnetic field \cite{tokura2006critical}, the position of the peak in the R(T) measurements of our devices does not change with magnetic field, indicating clearly that the magnetoresistance does not originate from the \ce{LSMO} layers \cite{galceran2016tunneling}. Altogether, the conductance of these devices is controlled mostly by the magnetic polarization of the barrier with respect to the metallic and ferromagnetic electrodes.

\begin{figure}
\center
\includegraphics[scale=0.45]{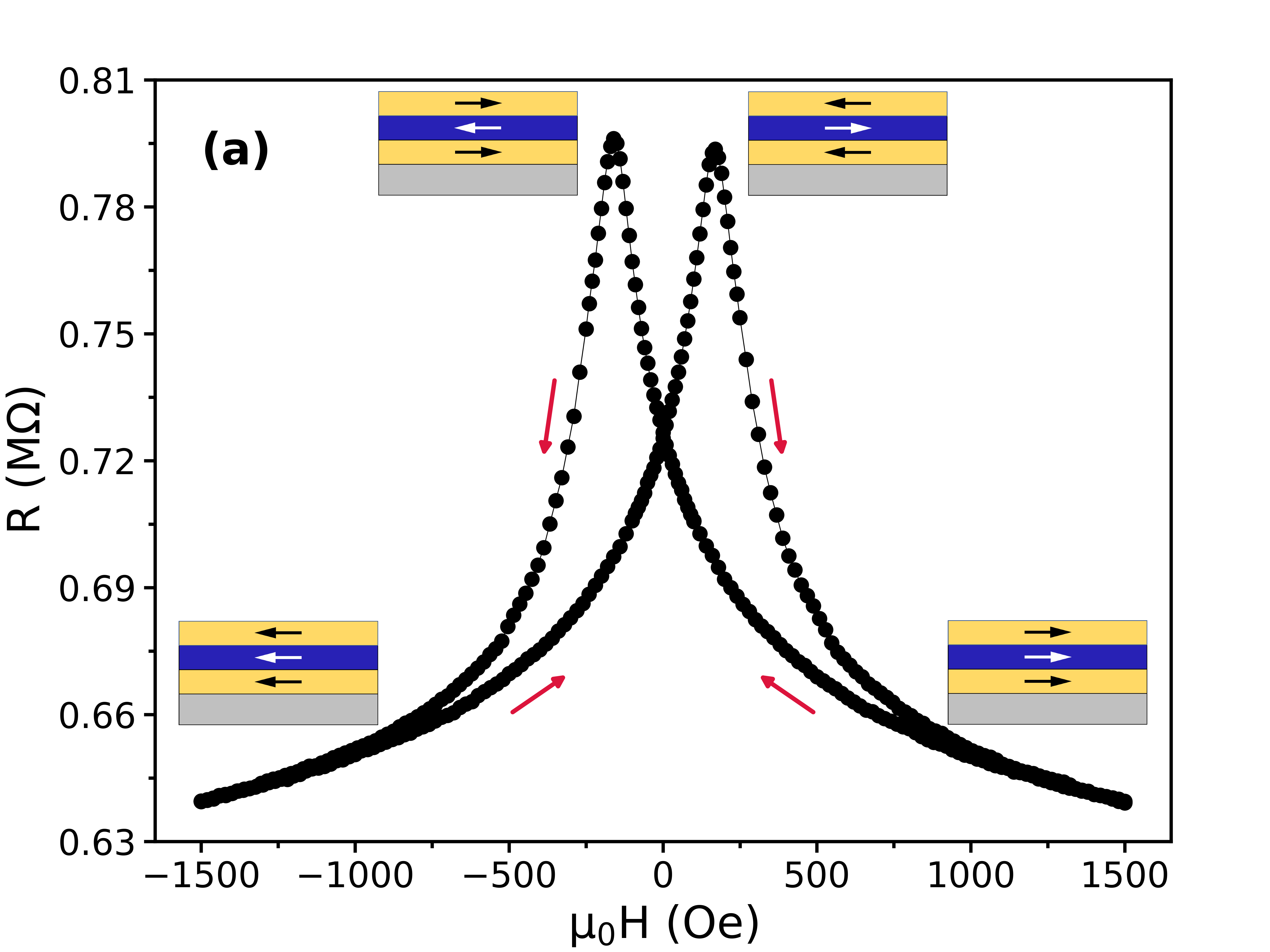}
\includegraphics[scale=0.45]{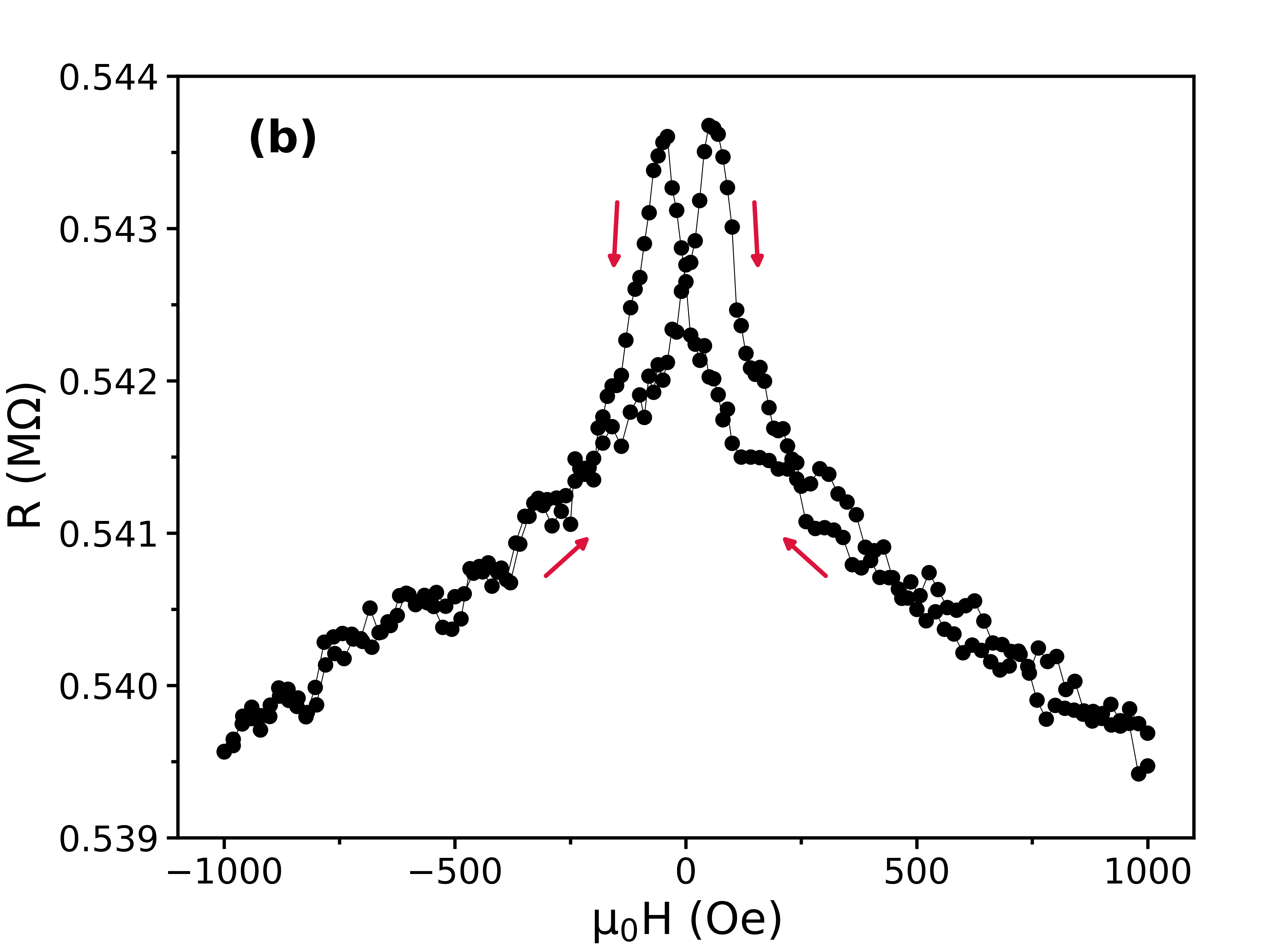}
\caption{TMR versus applied magnetic field at (a) \SI{10}{\kelvin} and (b) \SI{250}{\kelvin}. Fig.~(b) demonstrates that the tunneling process persists even above the apparent  transition temperature of the \ce{LNMO} barrier at \SI{180}{\kelvin}.}
\label{RvsH}
\end{figure}

\begin{figure}
\center
\includegraphics[scale=0.45]{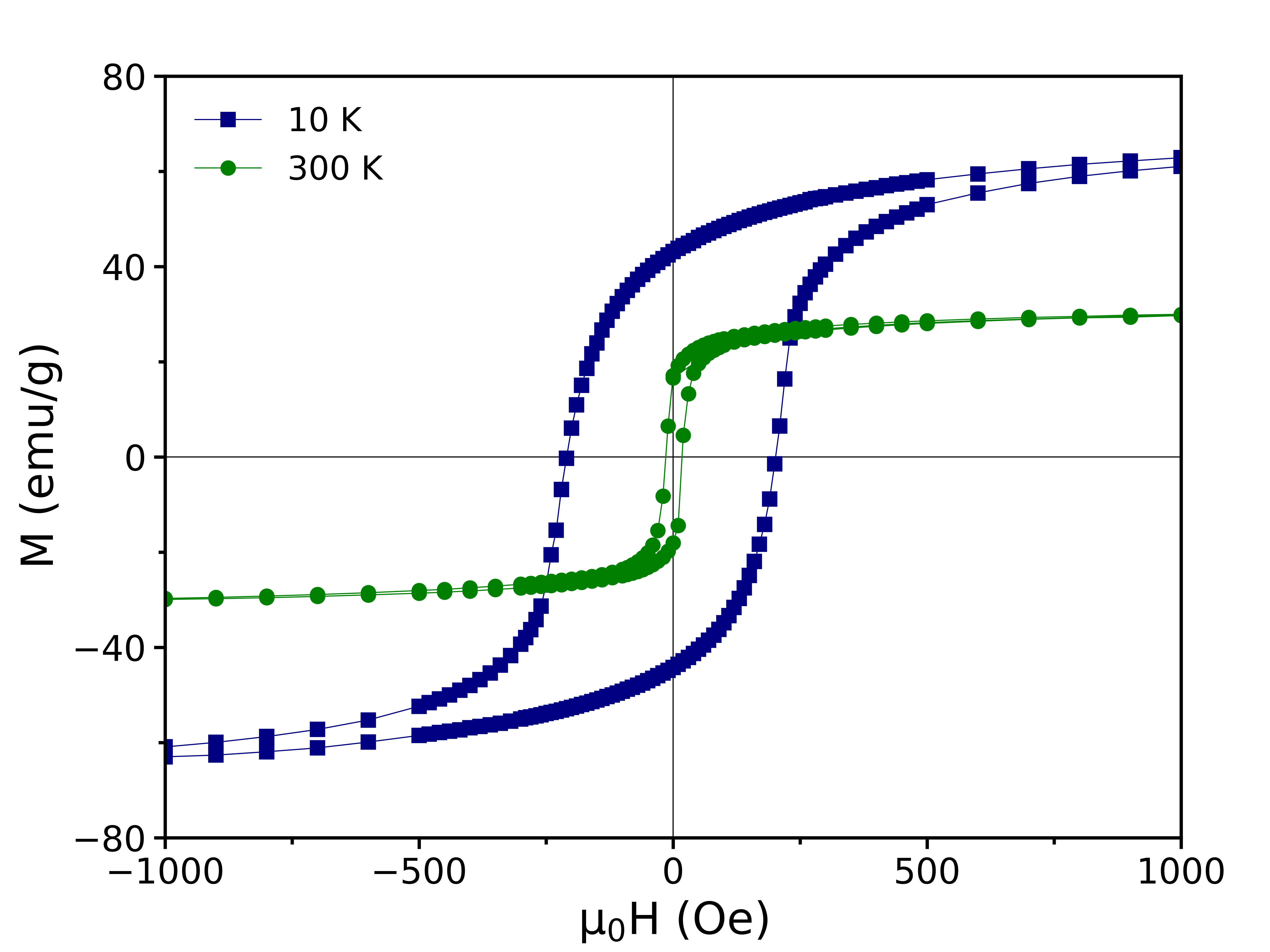}
\caption{Magnetic hysteresis loops of the MTJ device with magnetic field applied parallel to the surface of the sample at \SIlist[list-units=single]{10;300}{\kelvin}. The M(H) loop at \SI{300}{\kelvin} indicates that the \ce{LSMO} electrodes are still ferromagnetic at room temperature.}
\label{MvsHat10K}
\end{figure}

Figures~\ref{RvsH}~(a) and (b) present the magnetic field dependence of the MTJ resistance at \SIlist{10;250}{\kelvin} with the magnetic field applied parallel to the surface of the film. The junction shows a symmetric magnetoresistance hysteresis loop with low and high resistance states defined by the parallel and antiparallel alignment of the magnetic moments in \ce{LSMO} and \ce{LNMO} layers with respect to each other. This trend persists up to \SI{280}{\kelvin} (Fig.~\ref{RvsH}~(b) and also shown in supplemental material). As schematically illustrated in Fig.~\ref{RvsH}~(a), switching between the low and high resistance states is governed by the magnetization direction in the \ce{LNMO} barrier. In general, we have observed that \ce{LNMO} thin films present larger coercive fields ($\sim~$\SIrange[range-phrase=--]{500}{1000}{Oe}) than \ce{LSMO} films ($\sim~$\SIrange[range-phrase=--]{50}{300}{Oe}): see supplemental material. While \ce{LSMO} films show usually sharp polarization switches at their coercive field, \ce{LNMO} films tend to have broader polarity transitions (see supplemental material). In a sufficiently high magnetic field, the magnetic moments of all three layers are aligned and the junction stays in the lowest resistance state. With decreasing magnetic field, the magnetic moments of the \ce{LNMO} layer start flipping gradually and orient antiparallel to those of the \ce{LSMO} electrodes. This antiparallel configuration in some areas of the junction blocks the low-resistance conduction paths and consequently conduction occurs via another channel with higher resistance. The gradual increase of resistance continues as more magnetic domains flip in the \ce{LNMO} layer, until the magnetic field reaches the coercive field of \ce{LSMO}. At this point, the junction reaches its maximum resistance at \SI{+-160}{Oe}, where a large proportion of the magnetic domains in the \ce{LNMO} barrier are aligned in opposite direction with respect to those of the \ce{LSMO} electrodes. This magnetic field is very close to the coercive field of \ce{LSMO} (see supplemental material) and far from the coercive field of the device (see Fig.~\ref{MvsHat10K} and discussion below). From \SI{160}{Oe}, further increasing the magnetic field flips rapidly the magnetic moments of both \ce{LSMO} electrodes. If we assume that the top and bottom \ce{LSMO} electrodes in our devices have identical coercive fields, this rapid flip of both \ce{LSMO} electrodes results again in the all-parallel configuration and reestablishes the high conduction paths. Consequently, the resistance decreases with field above \SI{160}{Oe}. Unlike typical TMJs with a sharp switching between two resistance states at the different coercive fields of the ferromagnetic electrodes \cite{jo2000very, ishii2006improved}, the rounded shape of the MR peak in our device can be attributed to the presence of a magnetic spacer with a gradual switching of its magnetic moments.

\begin{figure}
\center
\includegraphics[scale=0.33]{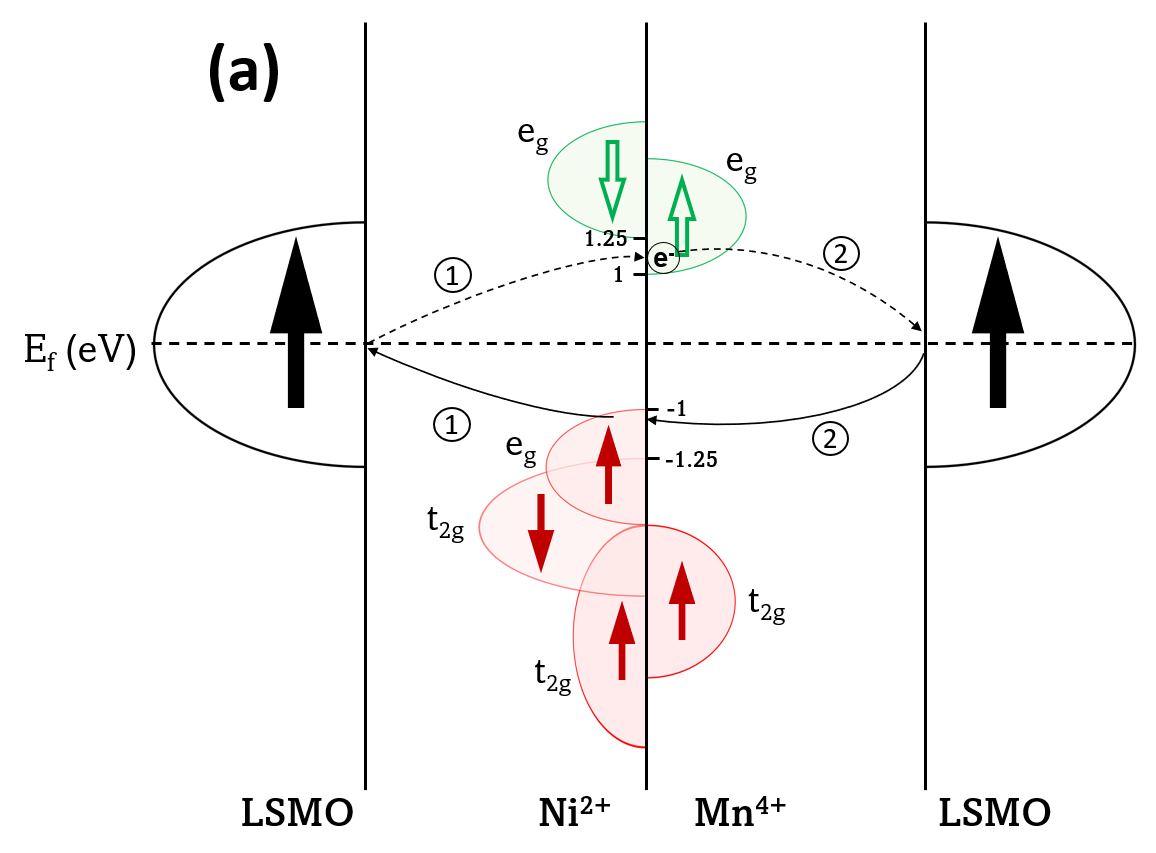}
\includegraphics[scale=0.33]{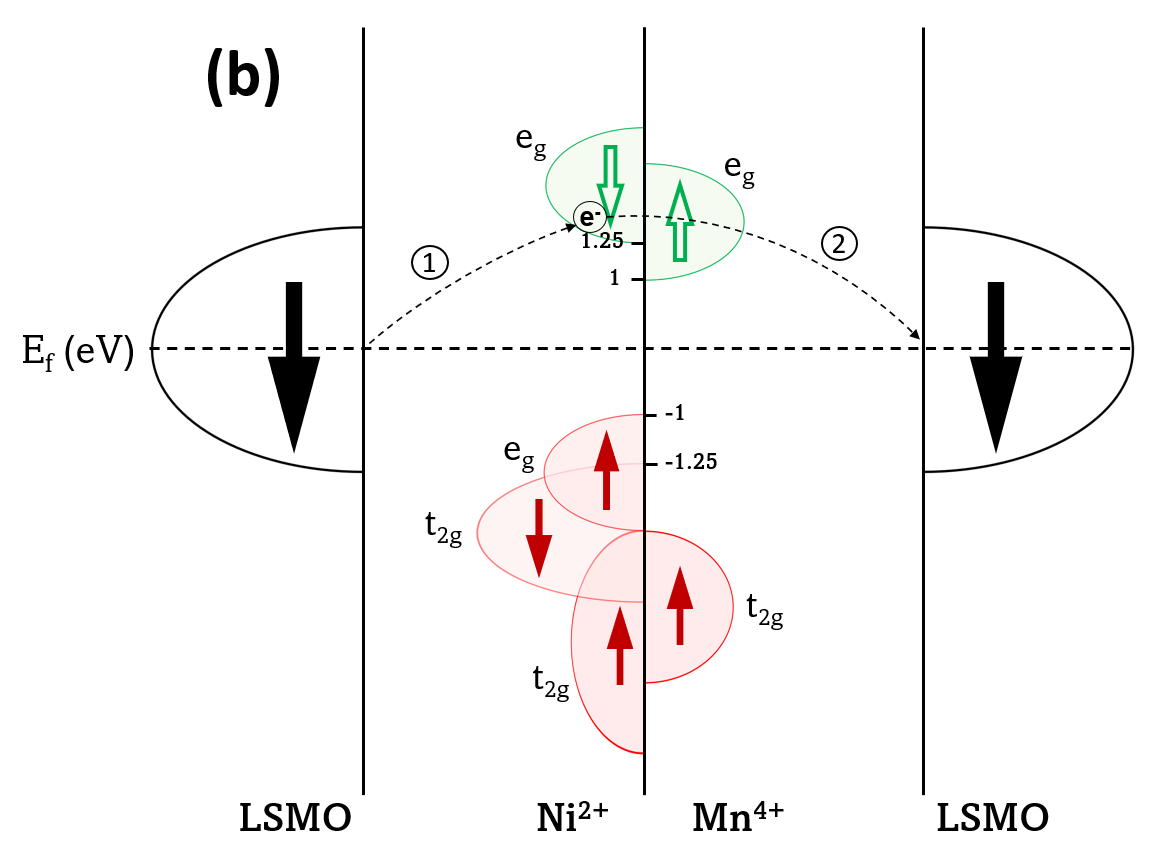}
\caption{Schematics of the conduction mechanism in our device in (a) parallel and (b) antiparallel configurations. Oxygen levels have been left out intentionally as they are not affecting the final outcome. The electron and hole channels are represented with the dashed and solid lines, respectively. 1 and 2 indicate the first and second hops during the tunneling process via the electron and hole channels.}
\label{R_mechanism}
\end{figure}

The difference in resistance between the two magnetic configurations in our MTJ devices originates from the contribution of different tunneling processes taking into account the location in energy of the spin-polarized occupied and unoccupied levels in \ce{LNMO}. Based on band structure calculations \cite{gauvin2018electronic}, these levels in \ce{LNMO} can be positioned roughly according to the schematics presented in Figure~\ref{R_mechanism} assuming that the Fermi energy ($E_{\mathrm{F}}$) of insulating \ce{LNMO} is sitting in the middle of its gap. Tunneling can take place through several channels. The first type of channels involves electrons in \ce{LSMO} $e_{g}$ levels accessing the empty $e_{g}$ levels above the Fermi energy of \ce{LNMO} and then reaching the $e_{g}$ levels of the other \ce{LSMO} electrode. The second type of channels involves instead hole tunneling through the occupied states of \ce{LNMO} below the Fermi energy. Since the electron spin polarization should be preserved during these different tunneling processes, the magnetic polarization of the barrier will select specific channels. The energy barrier height will then be set by the energy position of these empty and filled $e_{g}$ states of \ce{LNMO} relative to $E_{\mathrm{F}}$.

In Fig.~\ref{R_mechanism}~(a) showing the all-parallel configuration, the empty $e_{g}$ levels of \ce{Mn^{4+}} and the filled $e_{g}$ ones of \ce{Ni^{2+}} contribute to the tunneling current defining a barrier height of roughly \SI{1}{eV} for both channels. When both \ce{LSMO} electrodes are anti-parallel with respect to that of \ce{LNMO} as in Fig.~\ref{R_mechanism}~(b), the tunneling occurs through the empty $e_{g}$ levels of \ce{Ni^{2+}} implying a barrier height of roughly \SI{1.25}{eV}. This higher barrier explains the higher resistance when the polarizations of \ce{LNMO} and the \ce{LSMO} electrodes are opposite. In Fig.~\ref{R_mechanism}, we assumed that electrons can only tunnel between bands carrying the same symmetry explaining the absence of a hole channel in the anti-parallel configuration in Fig.~\ref{R_mechanism}~(b). In such a case, a lower number of channels will also contribute to increase the resistance. If tunneling does not require that the wavefunction symmetry must be preserved but only the spin, an additional hole channel through the occupied $t_{2g}$ levels should also be considered. However, since it is further away from  $E_{\mathrm{F}}$ than the occupied $e_{g}$ levels for the hole channel of the parallel configuration in Fig.~\ref{R_mechanism}~(a), it would lead also to a higher resistance. We should underline that doping, for example via oxygen vacancies, will likely shift $E_{\mathrm{F}}$ in the gap of \ce{LNMO}. However, this shift will only change the magnitude of the barrier heights for all the channels considered but not their role in the tunneling processes leaving unchanged the mechanism described above.

Figure~\ref{MvsHat10K} displays the magnetic hysteresis loops of the MTJ device at \SIlist[list-units=single]{10;300}{\kelvin} with the magnetic field applied parallel to the surface of the sample. Magnetization at \SI{10}{\kelvin} reaches saturation at magnetic fields of the order of \SI{1000}{Oe}, matching closely the field required to reach resistance reversibility in the R(H) measurements at the same temperature. It confirms that the magnetic moments of \ce{LNMO} layer saturate and completely align with those of the \ce{LSMO} electrodes at high field. Moreover, the coercive field ($H_{\mathrm{c}}$) of the devices is found to be \SI{210}{Oe} which is closer to the coercive field of a \ce{LNMO} monolayer (\SI{240}{Oe}) than that of a \ce{LSMO} monolayer with $H_{\mathrm{c}}\sim~$\SI{140}{Oe} (see supplemental material). In fact, this confirms that the switching field of \SI{160}{Oe} observed in the R(H) data in Fig.~\ref{RvsH}~(a) is related to the polarization switching of \ce{LSMO}. Also, the M(H) loop at \SI{300}{\kelvin} shows a sharp polarization switching confirming that the device is still magnetic at room temperature and its magnetization originates only from the \ce{LSMO} electrodes. Generally, MTJs made with a regular insulator have the same coercive and switching fields. In our case, the higher coercive field in the magnetization measurements originates from the contribution of the \ce{LNMO} layer, whereas the switching field (the peaks) observed in the R(H) measurements is mostly a signature of the coercive field of the \ce{LSMO} layers. These findings are consistent with the behavior observed in a spin-filter device containing \ce{BiMnO_{3}} as a ferromagnetic insulating barrier \cite{gajek2005spin}, where a coercive field of \SI{460}{Oe} was measured from magnetization measurements while the switching field was as low as \SI{100}{Oe}, corresponding to the coercive field of the \ce{LSMO} electrode. 

The ratio of TMR is extracted from the R(H) data collected for different temperatures up to \SI{300}{\kelvin} (see supplemental material) and is displayed in Figure~\ref{TMR}. The maximum TMR ratio is \SI{24}{\percent} at \SI{10}{\kelvin}, while it drops rapidly to \SI{0.1}{\percent} at \SI{280}{\kelvin}. No irreversibility is observed above this temperature even though the \ce{LSMO} layers remain ferromagnetic. Typically, the critical temperature of MTJs is associated with spin polarization at the interface of the ferromagnetic electrodes and the barrier, which usually decays much faster with temperature than the bulk magnetization. For instance, the critical temperature for \ce{LSMO}-based junctions with non-magnetic barriers such as \ce{SrTiO_{3}}, \ce{LaAlO_{3}} and \ce{TiO_{2}} are found to be \SIlist[list-units=single]{260;280;300}{\kelvin}, respectively, while \ce{LSMO} electrodes are magnetic up to \SI{350}{\kelvin} \cite{garcia2004temperature}. Unlike MTJs with non-magnetic barriers, this temperature of our devices is not only controlled by the magnetic properties of the electrodes at their interfaces, it also depends on the magnetization in the barrier. Nevertheless, the effect of non-optimal magnetic properties due to oxygen vacancies at electrode/barrier interfaces on TMR cannot be ruled out \cite{viret1997low}.

Furthermore, contrary to expectations, the TMR does not disappear above the apparent transition temperature of the \ce{LNMO} barrier at \SI{180}{\kelvin} observed by magnetization in Fig.~\ref{MvsT}. In fact, we notice that the R(H) loops become noisy and show a small TMR above this temperature (see Fig.~\ref{RvsH}(b)). The observation of the TMR above the apparent $T_{\mathrm{c}}$ of \ce{LNMO} can be ascribed to the presence of some persisting magnetic domains in the \ce{LNMO} layer with a higher transition temperature, up to $\sim$\SI{280}{\kelvin}. These domains are probably originating from regions of the \ce{LNMO} film with a high level of cationic ordering \cite{singh2010multiferroic}. These magnetic domains form only a small fraction of the volume of the \ce{LNMO} layer, so they could not be detected in the M(T) measurements dominated by the \ce{LSMO} transition above \SI{200}{\kelvin} but can be noticed in the R(H) data. Finally, the switching field that approaches closely the coercive field of \ce{LSMO} electrodes is also plotted as a function of temperature in Fig.~\ref{TMR}, as well. One can see that it follows almost the same trend as the TMR and that it cannot be detected above \SI{280}{\kelvin}, emphasizing the absence of TMR above the ferromagnetic to paramagnetic transition temperature of \ce{LNMO}.

\begin{figure}
\center
\includegraphics[scale=0.38]{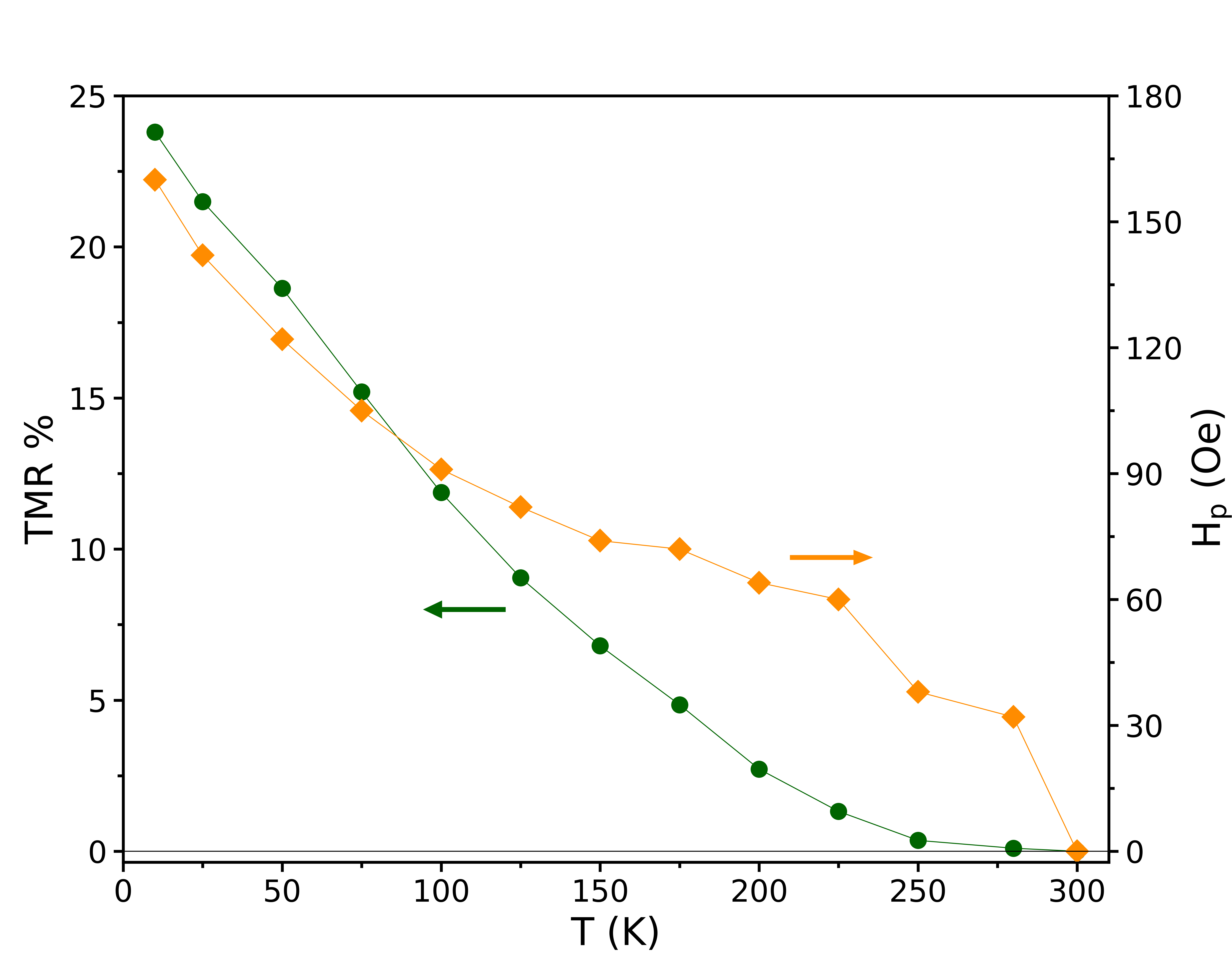}
\caption{left axis) The ratio of TMR in \ce{LSMO/LNMO/LSMO} junctions as a function of temperature. (right axis) The switching (peak) field that corresponds to the coercive field of \ce{LSMO} layers as a function of temperature.}
\label{TMR}
\end{figure}

Our results indicate that the use of \ce{LNMO} double perovskite as the barrier in MTJs can improve the operating temperature range of tunnel junctions containing a magnetic insulating barrier. Better performances are even expected if one can improve cationic ordering in \ce{LNMO}.

\section*{Conclusion}

In summary, we have presented a study of oxide-based magnetic tunnel junctions including a \ce{La_{2}NiMnO_{6}} (\ce{LNMO}) double perovskite as a ferromagnetic insulating barrier. The temperature dependence of the junction resistance shows a similar behavior to that of other oxide-based junctions with a peak in the mid-temperature range. We have measured a tunnel magnetoresistance of up to \SI{24}{\percent} at low temperature with a gradual switching between high and low resistance states. We demonstrated that the TMR depends on the direction of the relative polarity of the magnetization of both \ce{LSMO} electrodes and the \ce{LNMO} barrier with a mechanism involving the difference in barrier height driven by the location of the spin polarized empty and filled $e_{g}$ states in \ce{LNMO} around its Fermi energy. The junctions exhibit a TMR up to \SI{280}{\kelvin}, offering an improvement over existing spin-filtering junctions. A magnetic insulating barrier is an interesting path for achieving room temperature magnetoresistance  in oxide-based heterostructures.

\section*{Acknowledgment}

The authors thank B. Rivard, S. Pelletier and M. Dion for technical support. The authors also gratefully acknowledge Prof. I. Garate for the helpful discussions.  This work is supported by the Natural Sciences and Engineering Research Council of Canada (NSERC) under grant RGPIN-2018-06656, the Canada First Research Excellence Fund (CFREF), the Fonds de Recherche du Québec - Nature et Technologies (FRQNT) and the Université de Sherbrooke.

\bibliographystyle{nature}
\bibliography{M.Abbasi_MTJ.bib}

\begin{thebibliography}{10}

\bibitem{wolf2001spintronics}
Wolf, S., Awschalom, D., Buhrman, R., Daughton, J., von Moln{\'a}r, v.~S.,
  Roukes, M., Chtchelkanova, A.~Y., and Treger, D.
\newblock {\em
  \href{https://www.science.org/doi/10.1126/science.1065389}{Science}}{ \bf
  294}(5546), 1488--1495 (2001).

\bibitem{bratkovsky2008spintronic}
Bratkovsky, A.
\newblock {\em \href{https://doi.org/10.1088/0034-4885/71/2/026502}{Rep. Prog.
  Phys.}}{ \bf 71}(2), 026502 (2008).

\bibitem{julliere1975tunneling}
Julliere, M.
\newblock {\em \href{https://doi.org/10.1016/0375-9601(75)90174-7}{Phys. lett.,
  A}}{ \bf 54}(3), 225--226 (1975).

\bibitem{maekawa2002spin}
Maekawa, S. and Shinjo, T.
\newblock {\em Spin dependent transport in magnetic nanostructures}.
\newblock
  \href{https://www.routledge.com/Spin-Dependent-Transport-in-Magnetic-Nanostructures/author/p/book/9781420024579}{CRC
  press},  (2002).

\bibitem{bratkovsky1997tunneling}
Bratkovsky, A.
\newblock {\em \href{https://doi.org/10.1103/PhysRevB.56.2344}{Phys. Rev. B}}{
  \bf 56}(5), 2344 (1997).

\bibitem{lu1996large}
Lu, Y., Li, X., Gong, G., Xiao, G., Gupta, A., Lecoeur, P., Sun, J., Wang, Y.,
  and Dravid, V.
\newblock {\em \href{https://doi.org/10.1103/PhysRevB.54.R8357}{Phys. Rev. B}}{
  \bf 54}(12), R8357 (1996).

\bibitem{sun1997temperature}
Sun, J., Krusin-Elbaum, L., Duncombe, P., Gupta, A., and Laibowitz, R.
\newblock {\em \href{https://doi.org/10.1063/1.118651}{Appl. Phys. Lett.}}{ \bf
  70}(13), 1769--1771 (1997).

\bibitem{viret1997low}
Viret, M., Drouet, M., Nassar, J., Contour, J., Fermon, C., and Fert, A.
\newblock {\em \href{https://doi.org/10.1209/epl/i1997-00391-2}{EPL}}{ \bf
  39}(5), 545 (1997).

\bibitem{bowen2003nearly}
Bowen, M., Bibes, M., Barth{\'e}l{\'e}my, A., Contour, J.-P., Anane, A.,
  Lemaitre, Y., and Fert, A.
\newblock {\em \href{https://doi.org/10.1063/1.1534619}{Appl. Phys. Lett.}}{
  \bf 82}(2), 233--235 (2003).

\bibitem{meservey1994spin}
Meservey, R. and Tedrow, P.
\newblock {\em \href{https://doi.org/10.1016/0370-1573(94)90105-8}{Phys.
  Rep.}}{ \bf 238}(4), 173--243 (1994).

\bibitem{moodera1995large}
Moodera, J.~S., Kinder, L.~R., Wong, T.~M., and Meservey, R.
\newblock {\em \href{https://doi.org/10.1103/PhysRevLett.74.3273}{Phys. Rev.
  Lett.}}{ \bf 74}(16), 3273 (1995).

\bibitem{wei200780}
Wei, H., Qin, Q., Ma, M., Sharif, R., and Han, X.
\newblock {\em \href{https://doi.org/10.1063/1.2696590}{J. Appl. Phys.}}{ \bf
  101}(9), 09B501 (2007).

\bibitem{bowen2001large}
Bowen, M., Cros, V., Petroff, F., Fert, A., Martinez~Boubeta, C.,
  Costa-Kr{\"a}mer, J.~L., Anguita, J.~V., Cebollada, A., Briones, F.,
  De~Teresa, J., et~al.
\newblock {\em \href{https://doi.org/10.1063/1.1404125}{Appl. Phys. Lett.}}{
  \bf 79}(11), 1655--1657 (2001).

\bibitem{ikeda2008tunnel}
Ikeda, S., Hayakawa, J., Ashizawa, Y., Lee, Y., Miura, K., Hasegawa, H.,
  Tsunoda, M., Matsukura, F., and Ohno, H.
\newblock {\em \href{https://doi.org/10.1063/1.2976435}{Appl. Phys. Lett.}}{
  \bf 93}(8), 082508 (2008).

\bibitem{gajek2005spin}
Gajek, M., Bibes, M., Barth{\'e}l{\'e}my, A., Bouzehouane, K., Fusil, S.,
  Varela, M., Fontcuberta, J., and Fert, A.
\newblock {\em \href{https://doi.org/10.1103/PhysRevB.72.020406}{Phys. Rev.
  B}}{ \bf 72}(2), 020406 (2005).

\bibitem{luders2006spin}
L{\"u}ders, U., Bibes, M., Bouzehouane, K., Jacquet, E., Contour, J.-P., Fusil,
  S., Bobo, J.-F., Fontcuberta, J., Barth{\'e}l{\'e}my, A., and Fert, A.
\newblock {\em \href{https://doi.org/10.1063/1.2172647}{Appl. Phys. Lett.}}{
  \bf 88}(8), 082505 (2006).

\bibitem{leclair2002large}
LeClair, P., Ha, J., Swagten, H., Kohlhepp, J., Van~de Vin, C., and De~Jonge,
  W.
\newblock {\em \href{https://doi.org/10.1063/1.1436284}{Appl. Phys. Lett.}}{
  \bf 80}(4), 625--627 (2002).

\bibitem{kitamura2009ferromagnetic}
Kitamura, M., Ohkubo, I., Kubota, M., Matsumoto, Y., Koinuma, H., and Oshima,
  M.
\newblock {\em \href{https://doi.org/10.1063/1.3111436}{Appl. Phys. Lett.}}{
  \bf 94}(13), 132506 (2009).

\bibitem{jonsson2000reliability}
J{\"o}nsson-{\AA}kerman, B., Escudero, R., Leighton, C., Kim, S., Schuller,
  I.~K., and Rabson, D.
\newblock {\em \href{https://doi.org/10.1063/1.1310633}{Appl. Phys. Lett.}}{
  \bf 77}(12), 1870--1872 (2000).

\bibitem{xia2021angular}
Xia, H., Zhang, S., Li, H., Li, T., Liu, F., Zhang, W., Guo, W., Miao, T., Hu,
  W., Shen, J., et~al.
\newblock {\em \href{https://doi.org/10.1016/j.rinp.2021.103963}{Results
  Phys.}}{ \bf 22}, 103963 (2021).

\bibitem{galceran2016tunneling}
Galceran, R., Balcells, L., Pomar, A., Konstantinovi{\'c}, Z., Bagu{\'e}s, N.,
  Sandiumenge, F., and Mart{\'i}nez, B.
\newblock {\em \href{https://doi.org/10.1063/1.4946851}{AIP Advances}}{ \bf
  6}(4), 045305 (2016).

\bibitem{tokura2006critical}
Tokura, Y.
\newblock {\em \href{https://doi.org/10.1088/0034-4885/69/3/R06}{Rep. Prog.
  Phys.}}{ \bf 69}(3), 797 (2006).

\bibitem{jo2000very}
Jo, M.-H., Mathur, N., Todd, N., and Blamire, M.
\newblock {\em \href{https://doi.org/10.1103/PhysRevB.61.R14905}{Phys. Rev.
  B}}{ \bf 61}(22), R14905 (2000).

\bibitem{ishii2006improved}
Ishii, Y., Yamada, H., Sato, H., Akoh, H., Ogawa, Y., Kawasaki, M., and Tokura,
  Y.
\newblock {\em \href{https://doi.org/10.1063/1.2245442}{Appl. Phys. Lett.}}{
  \bf 89}(4), 042509 (2006).

\bibitem{gauvin2018electronic}
Gauvin-Ndiaye, C., Baker, T., Karan, P., Mass{\'e}, {\'E}., Balli, M., Brahiti,
  N., Eskandari, M., Fournier, P., Tremblay, A.-M., and Nourafkan, R.
\newblock {\em \href{https://doi.org/10.1103/PhysRevB.98.125132}{Phys. Rev.
  B}}{ \bf 98}(12), 125132 (2018).

\bibitem{garcia2004temperature}
Garcia, V., Bibes, M., Barth{\'e}l{\'e}my, A., Bowen, M., Jacquet, E., Contour,
  J.-P., and Fert, A.
\newblock {\em \href{https://doi.org/10.1103/PhysRevB.69.052403}{Phys. Rev.
  B}}{ \bf 69}(5), 052403 (2004).

\bibitem{singh2010multiferroic}
Singh, M., Truong, K., Jandl, S., and Fournier, P.
\newblock {\em \href{https://doi.org/10.1063/1.3362922}{J. Appl. Phys.}}{ \bf
  107}(9), 09D917 (2010).

\end{thebibliography}


\begin{figure*}
\center
{\Huge Supplemental material}
	
\end{figure*}

\begin{figure*}
\center
\includegraphics[scale=0.6]{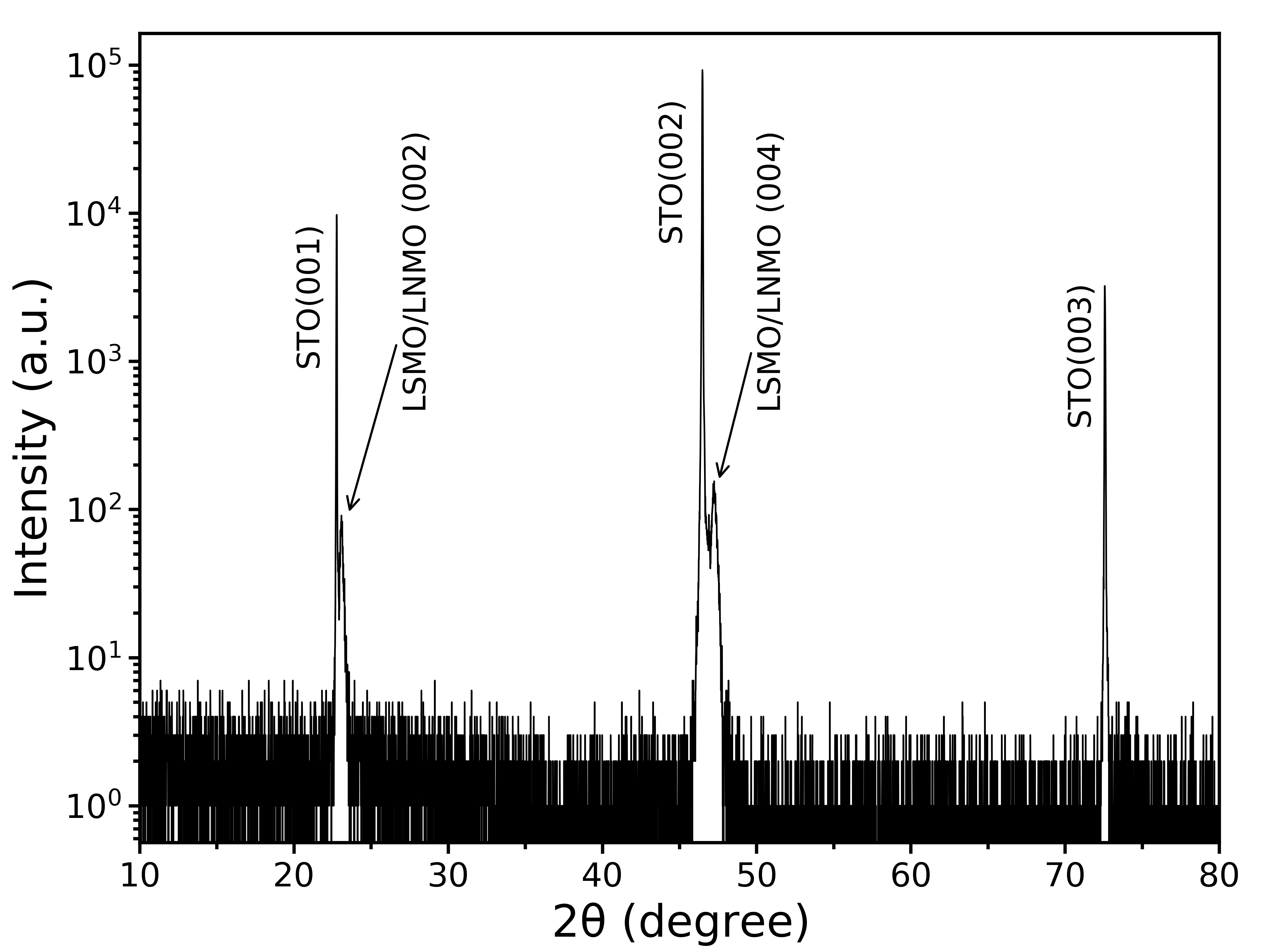}
\caption{X-ray diffraction pattern of a \ce{LSMO/LNMO/LSMO} MTJ from \ang{10} to \ang{80}. All peaks are assigned to the substrate, \ce{LSMO} and \ce{LNMO} layers.}
\end{figure*}

\begin{figure*}
\center
\includegraphics[scale=0.05]{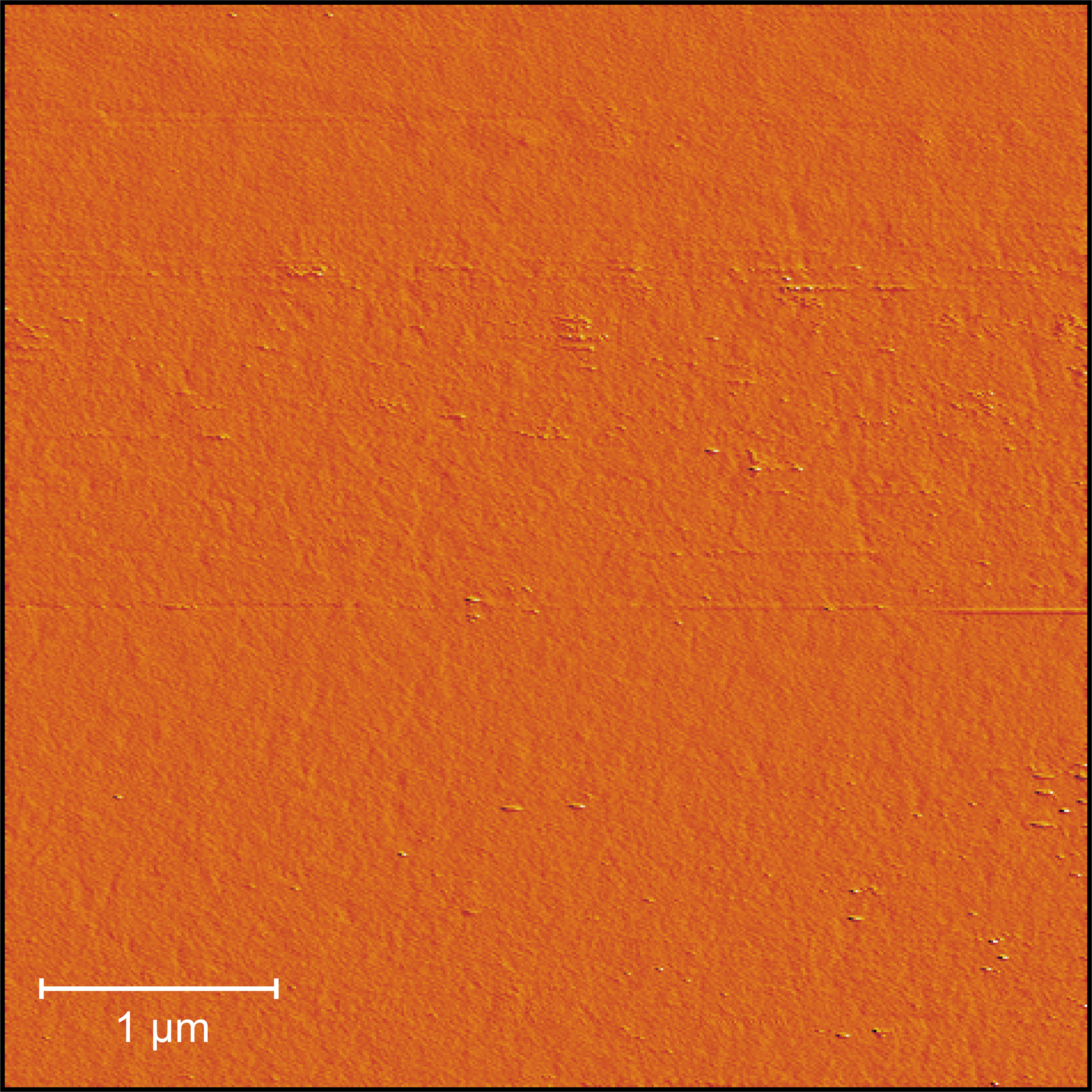}\\
\includegraphics[scale=0.6]{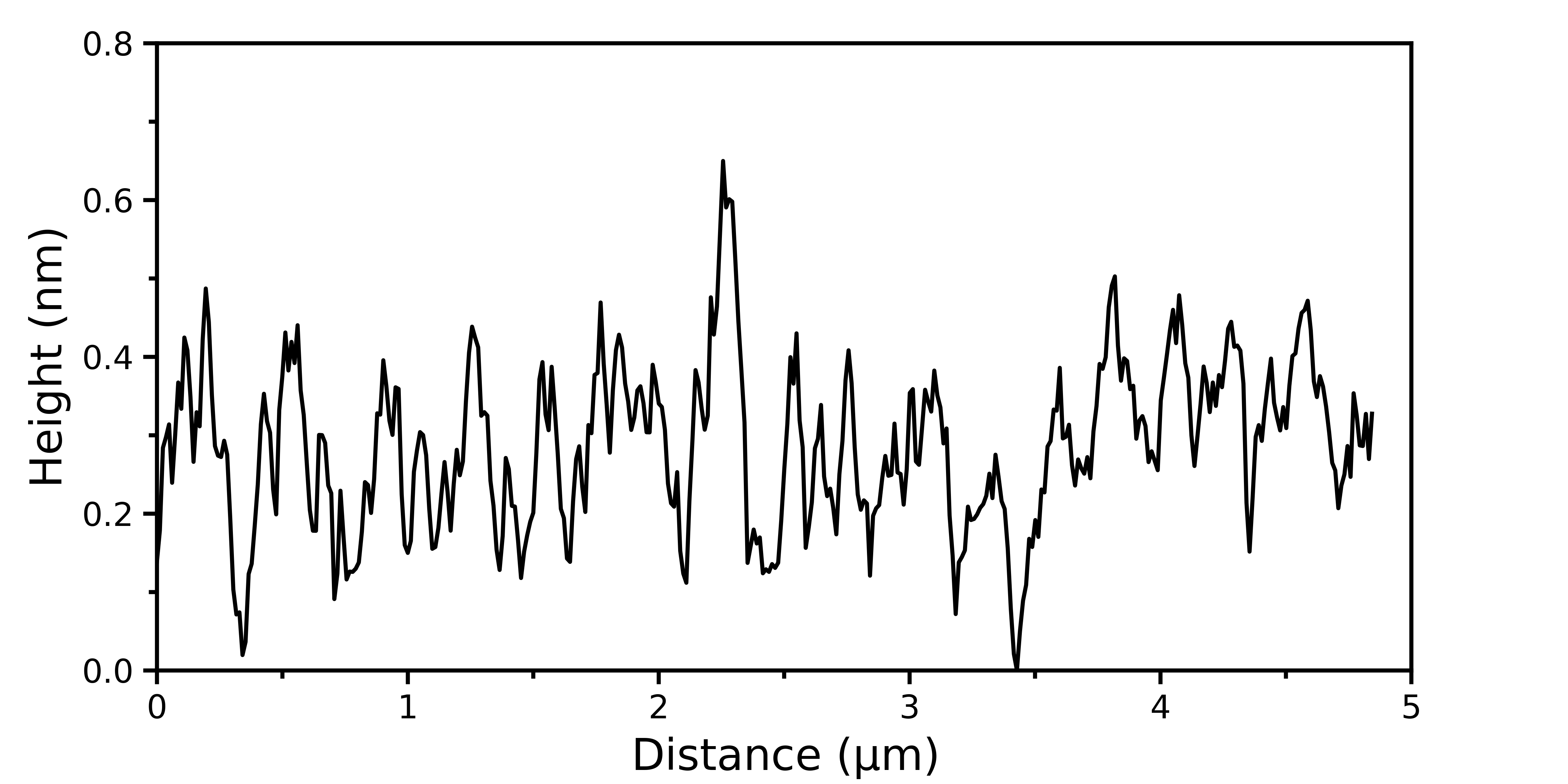}
\caption{Top: an AFM image of the bottom \ce{LSMO} layer in a MTJ before completing the device fabrication. Bottom: profile along a typical line indicating that the surface roughness of the layer is less than \SI{1}{\nm} over a lateral distance of \SI{5}{\mu m}. The surface roughness extracted from the top panel is less than \SI{1}{\nm} from Gwyddion.}
\end{figure*}

\begin{figure*}
\center
\includegraphics[scale=0.6]{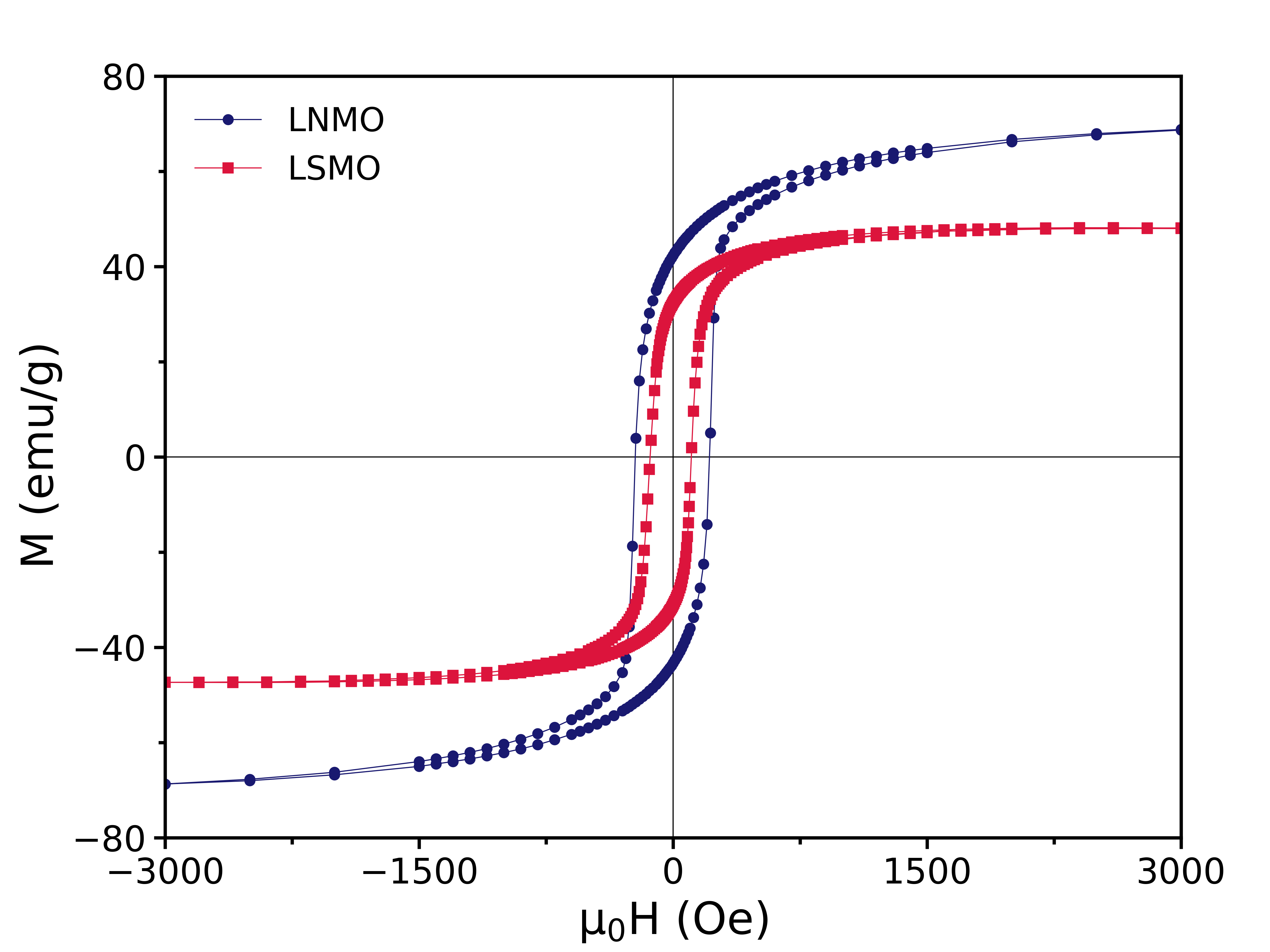}
\caption{Magnetic hysteresis loops of the \ce{LSMO} and \ce{LNMO} monolayers at \SI{10}{\kelvin}. As noticed, the \ce{LSMO} layer shows sharper polarization switches than the \ce{LNMO} layer and a lower coercive field ($H_{\mathrm{c}}$ is \SI{\pm 140}{Oe} for \ce{LSMO} and \SI{\pm 230}{\kelvin} for \ce{LNMO}).}
\end{figure*}

\begin{figure*}
	\center
	\begin{subfigure}{\textwidth}
	\center
	\includegraphics[width=0.8\textwidth]{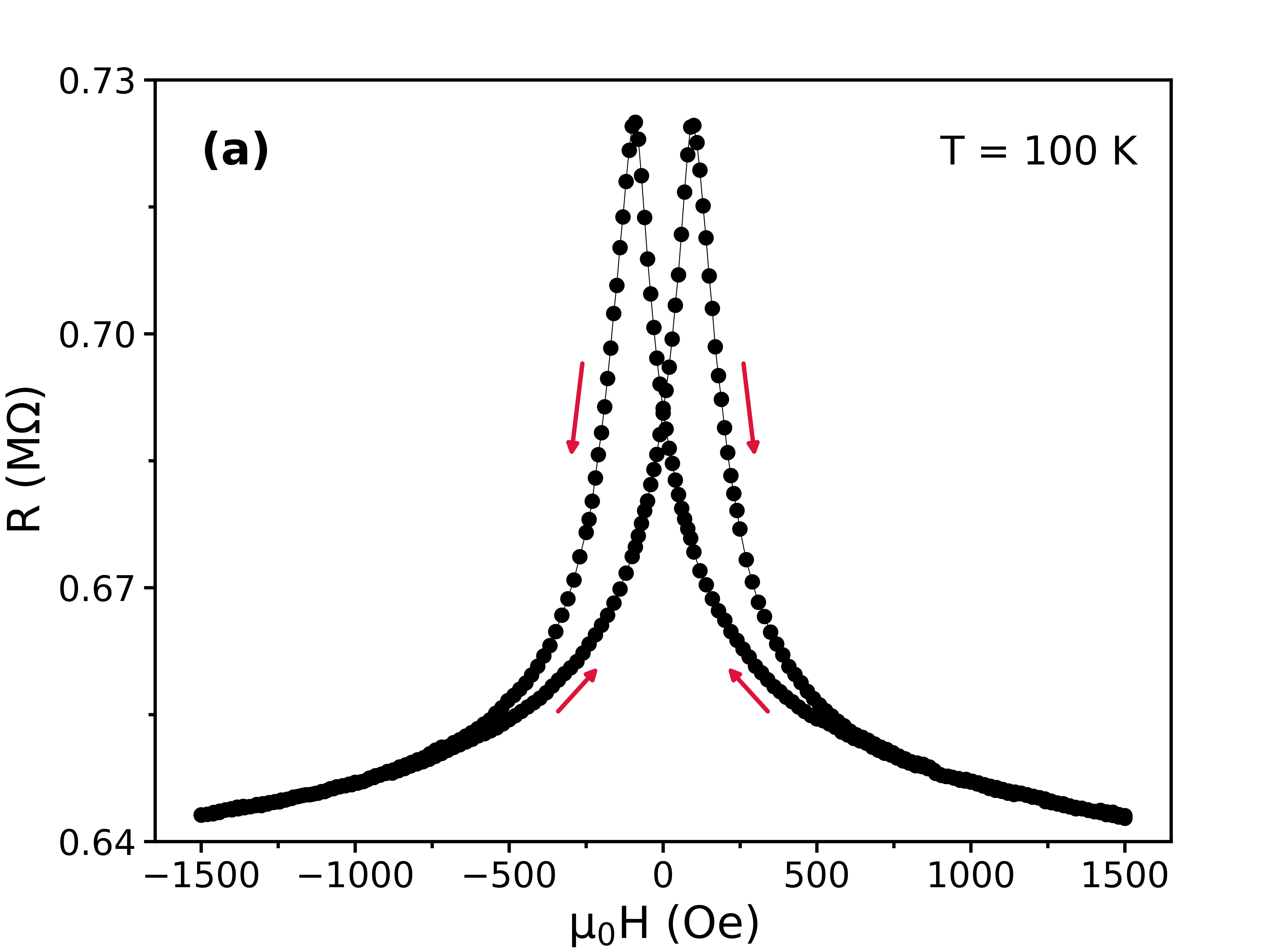}
	\end{subfigure}
\end{figure*}
\begin{figure*}
\ContinuedFloat
	\begin{subfigure}{\textwidth}
	\center
	\includegraphics[width=0.8\textwidth]{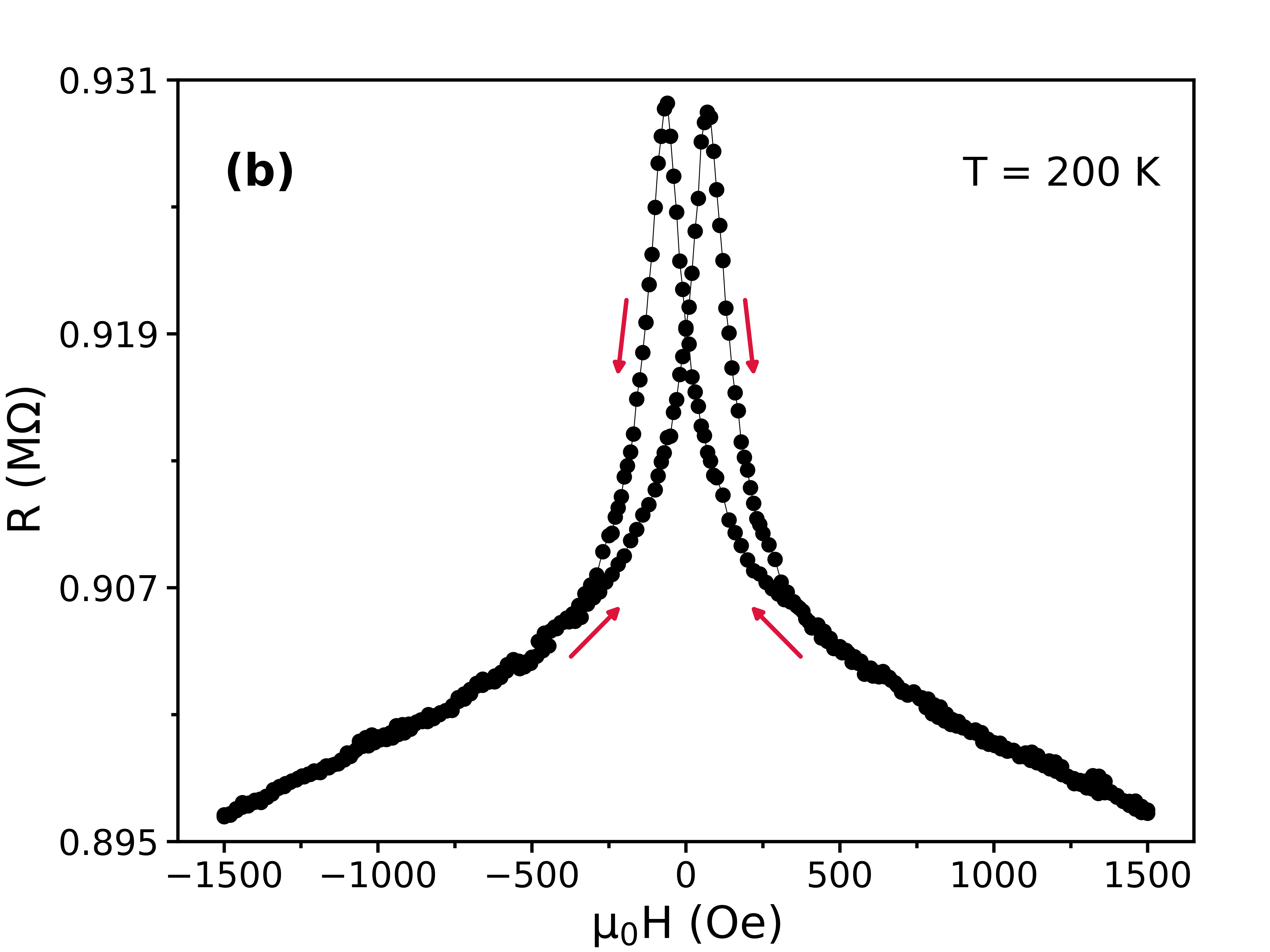}
	\end{subfigure}
\caption{Magnetic field dependence of the junction resistance at a) \SI{100}{\kelvin} and b) \SI{200}{\kelvin}.}
\end{figure*}

\end{document}